\documentclass[a4paper,mathleft]{an}
\usepackage{graphicx}
\usepackage{epsfig}
\usepackage{times}
\begin{document}


\title{Challenges and Opportunities for Helio- and Asteroseismology}

\author{W.~J.~Chaplin\thanks{\email{w.j.chaplin@bham.ac.uk}}}

\authorrunning{W. J. Chaplin}

\institute{School of Physics and Astronomy, University of Birmingham,
Edgbaston, Birmingham, B15 2TT, UK}

\received{XX Xxx 2010}
\accepted{XX Xxx 2010}
\publonline{later}

\keywords{stars: oscillations -- Sun: helioseismology}

\abstract{I consider some of the challenges and opportunities facing
helio- and asteroseismology, which reflect major themes of
presentation and discussion from the HELAS~IV international conference
``Seismological Challenges for Stellar Structure''. I concentrate in
particular on the exciting prospects for asteroseismology, now that
the field is being provided with data of unprecedented quality and in
unprecedented volumes.}

\maketitle

 \section{Introduction}
 \label{sec:intro}

The HELAS IV international conference ``Seismological Challenges for
Stellar Structure'' marked the successful conclusion of the
European-Commission Sixth Framework Programme (FP6) phase of the
European Helio- and Asteroseismology Network, ``HELAS''.  The
international helioseismology and asteroseismology communities have
benefited from HELAS in many significant ways: HELAS has supported a
diverse program of meetings and workshops, providing forums for
facilitating international collaboration, the exchange of knowledge
and expertise, and the development of new techniques for analyzing the
increasing volumes of high-quality data (with resulting benefits in
particular regarding provision of training of graduate students and
young scientists in the field).

Much has changed since the first HELAS international conference was
held at the University of Sheffield in 2006. Notable progress was then
being made in asteroseismology, thanks to ongoing programs of
episodic, ground-based telescope campaigns, observations by the MOST
microsatellite, and the successful exploitation of the WIRE
star-tracker data for asteroseismic studies. In helioseismology,
important developments continued thanks to data being collected by the
suite of helioseismic instruments on the ESA/NASA SoHo spacecraft, and
by the ground-based Birmingham Solar-Oscillations Network (BiSON).
CoRoT had not yet been launched, although that event was only a few
months away. The asteroseismology programme of the NASA Kepler mission
-- what would become the Kepler Asteroseismology Investigation (KAI)
-- was still in the early stages of being formulated. And the Sun was
heading towards its next solar minimum, with no unusual behaviour
forecast for that upcoming period.

Fast-forward to 2010, and the picture is much altered. CoRoT (Michel
et al. 2008) and Kepler (Gilliland et al. 2010) are now providing
stellar photometric data of unprecedented quality and length for
asteroseismology, and in unprecedented volumes. It is no
understatement to say that these missions are revolutionizing the
field. One of the highlights of the HELAS~IV conference was a special
session devoted to first asteroseismology results from Kepler,
spanning solar-like stars (Christensen-Dalsgaard et al. 2010; Chaplin
et al. 2010), red giants (Bedding et al. 2010), $\delta$-Scuti and
$\gamma$-Dor pulsators (Grigahc\`ene et al. 2010), and oscillations in
open clusters (Stello et al. 2010) and eclipsing binaries (Hekker et
al. 2010).

With regards to the Sun, there has of course been the unusually deep
and extended solar minimum between cycles 23 and 24, and
helioseismology has cast important light on this unusual
behaviour. The SoHo spacecraft is now almost at the end of its
operational life, and its helioseismic imaging capabilities have very
recently been superseded by the HMI instrument on board NASA's Solar
Dynamics Observatory (SDO), which was launched in 2010 February; while
the CNES PICARD satellite was launched in 2010 June.

My objective in this review is to comment on some of the opportunities
and challenges facing helio- and asteroseismology in the next few
years. I follow the tenor, emphasis and focus of presentation and
discourse at the HELAS~IV conference, and I have used results on
highlights from the meeting to illustrate and set the context of my
discussion. I have grouped the highlights into a few overarching
themes (reflecting of course my personal views on what were the main
themes of the meeting). Because I have had to be selective the list of
issues I present is by no means exhaustive, but hopefully provides a
fair reflection of the meeting.

I begin close to home, with the Sun, and the unusual recent solar
minimum.

\section{The Unusual Solar Minimum}
\label{sec:min}

The unusual behaviour during the current solar minimum of many
diagnostics and probes of solar activity has raised considerable
interest and debate in the scientific community (e.g., see the summary
by Sheeley 2010). The minimum has been unusually, and unexpectedly,
extended and deep. Polar magnetic fields have been very weak, and the
open flux has been diminished compared to other recent minima. 

Helioseismology has been used to probe the behaviour of sub-surface
flows during the solar minimum. Howe et al. (2009) found that the
equatorward progression of the lower branches of the so-called
torsional oscillations (east-west flows) was late in starting compared
to previous cycles. They flagged this delayed migration as a possible
pre-cursor of the delayed onset of cycle 24. The meridional
(north-south) flow also carries a signature of the solar cycle, which
converges toward the active-region latitudes and also intensifies in
strength as activity increases. Gonz\'alez-Hern\'andez (2010) found
that during the current minimum this component had developed to
detectable levels even before the visual onset of magnetic activity on
the solar surface.

The globally coherent acoustic properties of the recent solar minimum
have been studied extensively with low-degree p modes (by Broomhall et
al. 2009 and Salabert et al. 2009) and medium-degree p modes (Tripathy
et al. 2010). These studies have shown that while the surface proxies
of activity (e.g., the 10.7-cm radio flux) were quiescent and very
stable during the minimum, the p-mode frequencies showed much more
variability. Tripathy et al. (2010) noted further surprising behaviour
compared to the previous minimum, i.e., an apparent anti-correlation
of the p-mode frequency shifts and the surface proxies of activity,
during the epoch covered by the current minimum.

Broomhall et al. (2009) had suggested the possible presence of a
quasi-biennial modulation of the frequencies of the low-degree modes,
superimposed upon the well-established $\sim 11$-yr variation of the
frequencies. This has since been confirmed by further in-depth
analysis, which reveals a signature that is consistent in the
frequencies extracted from BiSON and GOLF data (Fletcher et al. 2010).
The fact that this biennial signature has similar amplitude in the
low-frequency \emph{and} the high-frequency modes used in this
analysis suggests that its origins lie deeper than the very
superficial layers responsible for the 11-year shifts.

With regards to the recent solar minimum, the challenge now is to try
to make sense of the unusual behaviour revealed by the helioseismic
data, and to follow the acoustic signatures and flows as the Sun
emerges from the minimum.  An important avenue of investigation will
be to compare results with those from the previous minimum, e.g., by
direct analysis of the differences of the frequencies observed at the
two epochs (Basu et al. 2010).

\section{Ensemble Seismology}
\label{sec:ensemble}

Kepler and CoRoT are providing exciting new opportunities to conduct
\emph{ensemble seismology} thanks to the large number of stars being
observed. (In the spirit of the Greek \emph{asteroseismology}, we
might be tempted to instead adopt the phrase
\emph{synasteroseismology}.) The prospects for solar-like stars are
particularly exciting. The large, homogeneous Kepler ensemble will for
the first time allow a proper seismic survey of a population of
solar-type field stars to be made. A statistical survey of trends in
important seismic parameters will allow tests of basic scaling
relations, comparisons with trends predicted from modelling, and lead
to important insights on the detailed modeling of stars. This work is
now in progress.

One may also pick from a large ensemble pairs, small groups or
sequences of stars that share common stellar properties, e.g., mass,
composition, or surface gravity. This opens the possibility to perform
what we might call \emph{differential} (or \emph{comparative})
seismology of stars, e.g., in analyzing the selected stars one may
eliminate or suppress any dependence of the modelling or results on
the common property, or properties. By selecting, for example, a
sequence of stars of very similar mass and composition it will be
possible to produce an exceedingly accurate and robust relative age
calibration, and give the potential to map evolutionary sequences of
internal properties and structures, allowing exquisite tests of
stellar evolutionary models.  By selecting stars with very similar
surface gravities, one may potentially probe differences in
near-surface physics and convection.

Further inferences on near-surface physics may potentially be acquired
by differential seismic analyses of simultaneous observations of the
same star made in photometry and Doppler velocity. For example, there
is the exciting prospect of our having simultaneous data on the
solar-type binary 16~Cyg from observations made in photometry by
Kepler and in Doppler velocity by the Stellar Observations Network
Group (SONG) (Grundahl et al. 2009). The opportunities for collecting
simultaneous data in photometry and Doppler velocity may be limited in
the near future to only a small number of stars, and as such it also
behoves us to look for twins of Kepler and CoRoT stars in data already
collected by ground-based telescopes.

Finally, even when stars do not share common properties, comparative
seismology still has the potential to aid accurate mode
identification, without which inferences drawn from the seismic data
would be greatly limited (e.g., see Bedding \& Kjeldsen 2010).

Notable successes for CoRoT have come from asteroseismic studies of
red giants. Data on large numbers of stars observed in the exoplanet
channel reveal ubiquitous signatures of red-giant oscillations.
Studies of this ensemble have shown clear observational evidence for
new seismic scaling relations (e.g., Hekker et al. 2009, Mosser et
al. 2010) predicted by theory (Stello et al. 2009a). The studies are
being extended to make inference on the red-giant population -- which
is dominated for the CoRoT sample by red-clump stars -- and to thereby
use the ensemble to test population synthesis models of the evolution
of the galaxy (Miglio et al. 2009).

CoRoT also revealed clear, unambiguous evidence for non-radial modes
in many of the red-giant stars (De~Ridder et al. 2009). This general
property has been confirmed and inferences from it extended by Kepler
observations of red giants. Bedding et al. (2010) reported results on
low-luminosity (H-shell burning) red giants, exploiting the ensemble
properties of the stars by presenting the seismic spectra in so-called
scaled, and semi-scaled, \'echelle diagrams (Bedding \& Kjeldsen
2010). These diagrams reveal clear ridges due to radial \emph{and}
non-radial modes, and show that for these stars the displacements in
frequency of the $l=1$ modes from the midpoints of the adjacent $l=0$
modes are negative in sign, unlike the positive displacements seen in
many main-sequence stars; moreover, the large width in frequency of
the $l=1$ ridge is likely due to mixed modes (see Montalb\'an et
al. (2010) for theoretical discussion of both issues).

Observations of classical $\delta$-Scuti and $\gamma$-Dor pulsators
have provided further opportunities for ensemble studies by
Kepler. Analysis revealed that nearly all members of the observed
ensemble were hybrid pulsators, showing both types of oscillation
(Grigahc\`ene et al. 2010). This unexpected result has implications
for how the stars are classified, and also for the theoretical
descriptions of the excitation of the modes.

Another obvious application of ensemble seismology is in the study of
oscillating stars in clusters. These stars of course share a common
history and, to within small uncertainties, may be regarded as lying
at the same distance. Stello et al. (2010) reported on oscillations of
red giants observed in one of the open clusters (NGC6819) in the
Kepler field. They found that the asteroseismic parameters could be
used to test cluster membership of the stars, i.e., by comparing the
observed frequency of maximum mode power with the frequency predicted
from the locations of the stars in the cluster colour-magnitude
diagram. In spite of having only limited data to hand when the study
was made, it was already possible to identify four possible
non-members despite those stars having a better than 80\,\% membership
probability from radial velocity measurements.

Opportunities also exist for using ground-based observations to
perform ensemble seismology of clusters. Saesen et al. (2010) have
studied classical pulsations of $\beta$\,Cephei stars in the open
cluster NGC6910. When cluster members are ordered by intrinsic
brightness, a clear systematic progression of the detected oscillation
periods is revealed, just like that seen in the NGC6819 red
giants. This suggests that simple inspection of the oscillation
spectra of the ensemble of classical pulsators may be used to diagnose
cluster membership, like the oscillation spectra of solar-like
pulsators in Stello et al. (2010).

There are also opportunities for applying ensemble approaches in
helioseismic studies, an obvious example being in the use of local
helioseismology data products. Komm \& Hill (2009) used local
helioseismology to perform an ensemble study of 1009 active regions.
They measured the subsurface flow properties beneath the regions, and
a statistical analysis of the results showed that the inferred
vorticity of the flows could help to distinguish flaring and
non-flaring active regions, discrimination that would not be possible
using, for example, X-ray flux data. Ensemble local helioseismology
has also been used to study differences (again, in a statistical
sense) in the properties found beneath active and quiet-Sun regions
(e.g., see Komm, Howe \& Hill 2009).

\section{New Data -- New Diagnostics, New Challenges}
\label{sec:diag}

With new data of unprecedented quality and length come new challenges
for data analysis, and new opportunities for making accurate and
precise inference on the stellar properties that were not possible
with older, inferior data products. There is the potential to develop
and apply new diagnostic data analysis tools to ensure that
contemporary data are exploited to their full potential. However,
incumbent on the scientist in any such endeavour is to understand the
limitations and biases inherent in those tools, dependent on the
quality of data that is available.

New asteroseismic data from the likes of Kepler and CoRoT, and new
helioseismic data from the likes of SDO and PICARD, will naturally
drive developments in data analysis. A useful strategy to help in
development and testing of analysis tools is to apply those tools to
artificial data. The data must of course be realistic, and due care
must be taken when applying conclusions drawn to results on real
data. Nevertheless, tests like these can be extremely instructive --
even vital -- in aiding, and steering appropriately, the development
of new tools; and for avoiding over-interpretation of results on real
data.

Inference from local helioseismology requires intensive work on
modelling of the propagation of acoustic waves in the presence of
magnetic fields. Realistic numerical simulations play a key r\^ole in
these studies (e.g., see Schunker (2010) and references
therein). Cameron et al. (2010) made numerical simulations of wave
propagation through cylindrically symmetric model sunspots, and found
good qualitative agreement between the modelled seismic signature and
the observed signature of the sunspot in active region NOAA~9787.

In asteroseismology, the asteroFLAG collaboration (Chaplin et
al. 2008) has been very active developing tools for application to
solar-like oscillators. A particular focus has been to develop tools
for automated analysis of many stars to meet the demands of the
Kepler survey phase, which will ultimately provide data on more than
1000 solar-like stars. The first application of some of these tools to
three bright, solar-like Kepler targets was reported in Chaplin et
al. (2010).  The collaboration has also tested methods for estimating
radii of solar-like stars (Stello et al. 2009b), using artificial
Kepler-like data.  The main goal of the Kepler Mission is to
characterize extrasolar planetary systems, particularly Earth-like
planets in the habitable zones of their host stars, and accurate and
precise stellar radii are required to constrain the sizes of the
detected exoplanets. Christensen-Dalsgaard et al. (2010) report the
first application of asteroseismology to known planet-hosting stars in
the Kepler field.

CoRoT data on F-type solar-like oscillators have challenged
pre-conceived ideas of how asteroseismic data are analysed
(Appourchaux et al. 2008; Barban et al. 2009; Garc\'ia et al. 2009;
Mosser et al. 2009). The large mode linewidths uncovered in these
stars forced modifications to ``peak bagging'' strategies for
extracting estimates of the mode parameters, and have posed challenges
for angular degree identification. Tests on artificial data are a
vital pre-requisite for improving the analysis applied to these stars
(e.g., see Benomar et al. 2009).

Experiments with artificial data on rapidly rotating intermediate and
high-mass stars have proven particularly instructive for helping to
understand the appearance of the observed classical oscillation
spectra in the presence of rapid rotation (Ballot et al. 2010a). The
impact of rapid rotation also cannot be ignored in some solar-like
oscillators, i.e., in young F-type stars. Here, modelling of the
seismic spectra must also account for second-order terms, which are
usually neglected in any analysis (Ballot 2010b).

Regions of stellar interiors where the structure changes abruptly
perturb the frequencies or periods of the oscillations (e.g., see
Houdek \& Gough 2007).  The characteristics of the signatures imposed
on the oscillations depend on the properties and locations of the
regions of abrupt structural change.  When the regions lie well within
the mode cavities, the near regular spacing of modes in frequency (for
p modes) or period (for g modes) are given periodic
displacements. There are, for example, signatures left by the
ionization of helium in the near-surface layers of solar-like stars.
Measurement of these signatures allows tight constraints to be placed
on the helium abundance, something that would not otherwise be
possible in such cool stars (because the ionization temperatures are
too high to yield usable photospheric lines for spectroscopy). There
are also signatures left by the locations of convective boundaries. It
is therefore possible to pinpoint the lower boundaries of convective
envelopes in cool stars. These regions are believed to play a key
r\^ole in stellar dynamos, and so this information is of great
importance to stellar dynamo modelers.

The key to extracting the signatures in solar-type stars is to have
sufficient precision in estimates of the mode frequencies. Multi-month
datasets are required, and tests undertaken with artificial
asteroseismic data (or real Sun-as-a-star helioseismic data) are
essential for elucidating diagnostic limits under different data
quality scenarios (e.g., see the tests undertaken by Basu et al. 2004;
Ballot, Turck-Chi\'eze \& Garc\'ia 2004; and Verner, Elsworth \&
Chaplin 2006). Mazumdar \& Michel (2010) report attempts to extract
periodic signatures from CoRoT frequency data on the F-type solar-like
star HD49933.

Frequencies of classical pulsations may be measured to much higher
precision than their intrinsically damped, solar-like
cousins. However, only recently has the first evidence been presented
(by Degroote et al. 2010) for a periodic displacement of the g-mode
frequencies of a classical pulsator (the B3V star HD50230). It was the
almost unbroken sequence of multi-month observations made by the CoRoT
spacecraft that provided the data quality necessary to uncover the
signal in the g-mode periods. This signal is thought to be a signature
(e.g., due to mixing) from the boundary of the convective core and
radiative envelope.

When regions of abrupt structural change do not lie well within the
mode cavities, the signatures they leave are more subtle. This is the
case for the signatures left by small convective cores in solar-type
stars slightly more massive than the Sun (Cunha \& Metcalfe 2007). The
measurement of the frequency dependence of prominent frequency
separations of the low-degree modes provides a potential diagnostic of
both the presence and size of a convective core (Brand\~ao \& Cunha,
2010). The mixing implied by the convective cores, and the possibility
of mixing of fresh hydrogen fuel into the nuclear burning cores --
courtesy of the regions of overshoot -- affects the main-sequence
lifetimes. Tests on artificial seismic data are essential to
understand the prospects, and potential limitations, for inference on
convective cores.

Finally, we touch on the diagnostic potential of avoided crossings
(Osaki 1975; Aizenman et al. 1977) in solar-like stars.  These avoided
crossings are a tell-tale indicator that the stars have evolved
significantly. In young solar-type stars there is a clear distinction
between the frequency ranges that will support acoustic (pressure, or
p) modes and buoyancy (gravity, or g) modes.  As stars evolve, the
maximum buoyancy (Brunt-V\"ais\"al\"a) frequency increases. After
exhaustion of the central hydrogen, the buoyancy frequency in the deep
stellar interior may increase to such an extent that it extends into
the frequency range of the high-order acoustic modes. Interactions
between acoustic modes and buoyancy modes may then lead to a series of
avoided crossings, which affect (or ``bump'') the frequencies and also
change the intrinsic properties of the modes, with some taking on
mixed p and g characteristics.

Measurement of the these avoided crossings has the potential to
provide exquisite constraints on the fundamental stellar properties.
Very little data have been available historically. Observational
evidence for avoided crossings had been uncovered in ground-based
asteroseismic data on two bright stars, $\eta$\,Boo and $\beta$~Hyi
(Christensen-Dalsgaard et al. 1995; Kjeldsen et al. 1995; Bedding et
al. 2007; Do\u{g}an et al. 2009).  Deheuvels et al. (2009) have also
recently reported evidence for avoided crossings in CoRoT observations
of the G-type star HD49385, with Deheuvels \& Michel (2009) using an
elegant analysis on coupled oscillators to discuss the results.
Kepler now promises dramatic changes in this area, because within the
large Kepler ensemble of solar-like oscillators there is a clear
selection of stars showing avoided crossings.  One of the three
solar-like stars selected for the first Kepler paper on solar-like
oscillators -- KIC~11026764 -- is a beautiful case in point, and has
been the subject of further in-depth study (Metcalfe et al. 2010). The
data provided by Kepler will by necessity drive developments in
analysis to allow the diagnostic potential of avoided crossings to be
fully exploited.

\acknowledgements

The author would like to thank the SOC and LOC of the HELAS~IV
conference for their kind invitation to present the summary of the
meeting. He acknowledges financial support from both HELAS, and the UK
Science Technology and Facilities Council (STFC), and thanks
C.~Constantinidou and B.~A.~Miller for useful advice concerning Greek
naming.

\end{document}